\documentclass[prl,showpacs,twocolumn,amssymb,amsmath,superscriptaddress]{revtex4}
\usepackage{graphicx}

\begin{document}
\title{Stochastic polarization formation in exciton-polariton Bose-Einstein condensates}

 \author{D. Read}
 \affiliation{School of Physics and Astronomy, University of Southampton, Highfield, Southampton SO17 1BJ, UK}
 \author{P. J. Membry}
 \affiliation{School of Physics and Astronomy, University of Southampton, Highfield, Southampton SO17 1BJ, UK}
 \author{T. C. H. Liew}
 \affiliation{School of Physics and Astronomy, University of Southampton, Highfield, Southampton SO17 1BJ, UK}
 \author{Yuri G. Rubo}
 \affiliation{School of Physics and Astronomy, University of Southampton, Highfield, Southampton SO17 1BJ, UK}
 \affiliation{Centro de Investigaci\'on en Energ\'{\i}a, Universidad Nacional Aut\'onoma de M\'exico, Temixco, Morelos, 62580, Mexico}
 \author{A. V. Kavokin}
 \affiliation{School of Physics and Astronomy, University of Southampton, Highfield, Southampton SO17 1BJ, UK}
 \affiliation{Marie-Curie Chair of Excellence ``Polariton devices'', University of Rome II, 1, via della Ricerca Scientifica, Rome, 00133, Italy}

\date{July 4, 2008}

\begin{abstract}
We demonstrate theoretically the spontaneous formation of a
stochastic polarization in exciton-polariton Bose-Einstein
condensates in planar microcavities under pulsed excitation. Below
the threshold pumping intensity (dependent on the polariton
life-time) the average polarization degree is close to zero, whilst
above threshold the condensate acquires a polarization described by
a (pseudospin) vector with random orientation, in general. We
establish the link between second order coherence of the polariton
condensate and the distribution function of its polarization. We
examine also the mechanisms of polarization dephasing and
relaxation.
\end{abstract}

% 78.67.-n Optical properties of low-dimensional, mesoscopic, and nanoscale materials and structures
% 71.36.+c Polaritons (including photon-phonon and photon-magnon interactions)
% 42.25.Kb Coherence
% 42.55.Sa Microcavity and microdisk lasers

\maketitle

%%%%%%%%%%%%%%%%%%%%%%%%%%%%%%%%%%%%%%%%%%%%%%%%%%%%%%%%%%%%%%%%%

\emph{Introduction.}---Bose-Einstein condensation (BEC) of
exciton-polaritons in semiconductor microcavities has recently been
demonstrated experimentally~\cite{Kasprzak06,Balili07,Lai07}. This
important discovery opens the perspective for various applications
including room-temperature polariton lasers~\cite{Christopoulos07},
microscopic optical parametric oscillators~\cite{Diederichs06}, and
optical logic gates~\cite{Leyder07}. In this context, it is
important to establish a reliable experimental criterion for the BEC
of polaritons. From the point of view of the Landau theory of phase
transitions~\cite{Lifshitz}, BEC requires the build up of an order
parameter physically associated with the macroscopic wave-function
of the polariton condensate~\cite{Onsager}. The phase of the order
parameter is chosen by the system undergoing BEC spontaneously; the
spontaneous symmetry breaking~\cite{Goldstone} is considered as the
``smoking gun'' for BEC. The order parameter build up is also
accompanied by a decrease of the second order coherence parameter,
$g_2(0)$, indicating the change from a thermal distribution of
polaritons to a coherent distribution \cite{Laussy04}.
Unfortunately, due to the limited time resolution of the
Hanbury-Brown and Twiss experimental set up, it is very hard to
measure $g_2(0)$ with a good accuracy~\cite{Deng02,LeSiDang2008}.

In recent work~\cite{Laussy06,Shelykh06} it has been suggested that
the build-up of the order parameter can be evidenced by polarization
measurements. Effectively, the polarization degree of light emitted
by polariton condensates contains information on the amplitudes and
relative phases of both components of the spinor wave function of
the condensate. In this Letter we present the kinetic model of
spontaneous formation of the polarization vector during the course
of polariton BEC. Indeed, the polarization vector provides the same
information on the condensate (complex) order parameter, as the
second order coherence; this allows a straightforward method of
evidencing the existence of spontaneous symmetry breaking in a
polariton system under pulsed excitation.

Existing theories for the dynamics of polariton BEC have been based
either on the semiclassical Boltzmann equations describing the
energy relaxation of polaritons but neglecting the phase of the
condensate~\cite{Malpuech02}, or on the Gross-Pitaevskii equation
and its generalizations, assuming the existence of a coherent
condensate from the very
beginning~\cite{Carusotto04,Wouters07,GPgens}. The present model
bridges the gap between these two approaches and allows one to
describe the formation of a coherent polariton condensate from an
incoherent ensemble of polaritons.

The formation of a polariton condensate is stochastic in its nature
since polaritons entering the condensate have random phases and
polarizations. Above the stimulation threshold when the average
population of the lowest energy polariton state exceeds 1, the
stochastic phase and polarization of the condensate start being
amplified and stabilize due to the stimulated scattering of
polaritons from an incoherent reservoir. Polariton interactions with
acoustic phonons and between themselves make the dynamics of the
order parameter complex and non-trivial since they lead to dephasing
and polarization relaxation. Here we treat these effects within the
Landau-Khalatnikov approach~\cite{HHreview77} generalized to
describe a two-component interacting Bose system.

\emph{Formalism.}---We consider BEC into one spin-degenerate
ground-state level. The evolution of the order parameter
$\psi_\sigma(t)$ is described by the Langevin type equation:
\begin{equation}
 \label{LangevinEq}
 \frac{d\psi_\sigma}{dt}=\frac{1}{2}\left[W(t)-\Gamma_c\right]\psi_\sigma
 +\theta_\sigma(t)-\frac{\mathrm{i}}{\hbar}\frac{\delta H_{int}}{\delta\psi_\sigma^*}
 -\gamma\mathfrak{R}_\sigma.
\end{equation}
Here $\sigma=\pm 1$ denotes the two spin (circularly polarized)
components of the order parameter~\cite{Kirill}, $W(t)$ is the
income rate, $\Gamma_c^{-1}$ is the polariton lifetime in the
condensate, $\theta_\sigma(t)$ is the noise defined below. The third
term in the righthand side of Eq.~(\ref{LangevinEq}) describes the
spin-dependent polariton-polariton interactions in the condensate
\begin{equation}
 \label{Hint}
 H_{int}=\frac{1}{2}\left[\alpha_1(|\psi_{+1}|^4+|\psi_{-1}|^4)
                          +2\alpha_2|\psi_{+1}|^2|\psi_{-1}|^2\right],
\end{equation}
where $\alpha_1>0$ defines the repulsion energy of two polaritons
with the same spin and $\alpha_2<0$ defines the attraction energy of
two polaritons with opposite spins. Finally, the last term in
Eq.~(\ref{LangevinEq}) describes the relaxation of the order
parameter. The form of the functional
$\mathfrak{R}_\sigma[\{\psi_\sigma,\psi_\sigma^*\}]$ depends on the
mechanism of relaxation.

In the particular case of low condensate occupations, when one can
neglect the polariton-polariton interactions and the order parameter
relaxation, Eq.~(\ref{LangevinEq}) reduces to the kinetic equation
for the order parameter derived in~\cite{Rubo04} and originated from
the quantum kinetic equation describing the evolution of the
condensate density matrix~\cite{Rubo03}. In this case the total
intensity of the white noise, $\theta_\sigma(t)$, is given by the
income rate of polaritons into the condensate, $W(t)$. The
correlator of the noise is
\begin{equation}
 \label{NoiseCor}
 \left<\theta_\sigma(t)\theta^*_{\sigma^\prime}(t^\prime)\right> =
 \frac{1}{2}W(t)\delta_{\sigma\sigma^\prime}
 \delta(t-t^\prime).
\end{equation}
This noise is responsible for the phase and polarization
fluctuations in the ground state of the polariton system both below
and above the condensation threshold. Above threshold, when the
condensate occupation $n(t)=|\psi_{+1}|^2+|\psi_{-1}|^2$ is not
small, dephasing due to polariton self-interaction~\cite{Porras03}
becomes also important.

In general, the income rate $W(t)$ should be found by solving the
semiclassical Boltzmann equation for the polariton relaxation into
the condensate~\cite{Malpuech02}. In what follows, however, we adopt
a simple model~\cite{Wouters07} considering all the polaritons that
are not in condensate as a single incoherent reservoir. The
reservoir occupation number $N_r(t)$ satisfies the kinetic equation:
\begin{equation}
 \label{Reservoir}
 \frac{dN_r}{dt}=-\Gamma_rN_r-W(t)[n(t)+1]+P(t),
\end{equation}
where $P(t)$ is the incoherent pump rate and $\Gamma_r^{-1}$ is the
life-time of polaritons in reservoir (and usually
$\Gamma_r\ll\Gamma_c$). The exact dependence of the income rate on
the reservoir occupation is defined by the relaxation mechanism. In
the simplest case of polariton-phonon relaxation they are
proportional to each other, $W(t)=rN_r(t)$.

In the case of very short pulsed excitation, with pulse duration
much less then $\Gamma_c^{-1}$, the pump is reduced to the initial
condition $N_r(0)=\int P(t)dt$ for the reservoir concentration.
Eq.~(\ref{Reservoir}) with $P=0$ is then solved simultaneously with
Eq.~(\ref{LangevinEq}) considering that the condensate is not
initially populated, i.e., $\psi_\sigma(0)=0$. First, we present the
results for the case when the polarization relaxation can be
neglected ($\gamma=0$), and then we describe the effects of
relaxation of the order parameter.

%%%%%%%%%%%%%%%%%%%%%%%%%%%%%%%%%%%%%%%%%%%%%%%%%%%%%%%%%%%%%%%%%%%%%
\emph{Formation of the order parameter.}---The complex order
parameter of the condensate $\psi_\sigma$ cannot be observed
directly in a photoluminescence (PL) experiments. On the other hand,
the polarization resolved PL gives access to the components of the
condensate pseudospin (Stokes vector)~\cite{Kirill}. These
components are related to $\psi_\sigma$ as
$S_x=(1/2)(\psi_{+1}^*\psi_{-1}^{\vphantom{*}}+\psi_{+1}^{\vphantom{*}}\psi_{-1}^*)$,
$S_y=(\mathrm{i}/2)(\psi_{+1}^*\psi_{-1}^{\vphantom{*}}-\psi_{+1}^{\vphantom{*}}\psi_{-1}^*)$,
$S_z=(1/2)(|\psi_{+1}|^2-|\psi_{-1}|^2)$. Note that the condensate
occupation, $n$, can be calculated from $n^2=4(S_x^2+S_y^2+S_z^2)$.
Averages and statistics of different experimentally observable
quantities are presented in this subsection. In what follows we will
be interested mostly in time-integrated quantities which will be
denoted by a bar, e.g., $\int\psi_\sigma dt=\overline{\psi_\sigma}$.
The averaging over multiple pulses (i.e., over realizations of
noise) will be denoted by angular brackets as in
Eq.~(\ref{NoiseCor}). In numerical calculations we used the
parameters $\Gamma_c=0.5\,\mathrm{ps}^{-1}$,
$\Gamma_r/\Gamma_c=0.01$, $\alpha_1=1.8\,\mu\mathrm{eV}$,
$\alpha_2/\alpha_1=-0.1$, and
$r=10^{-4}\,\mathrm{ps}^{-1}$~\cite{NotePar}.

\begin{figure}[t]
\includegraphics[width=3.2in]{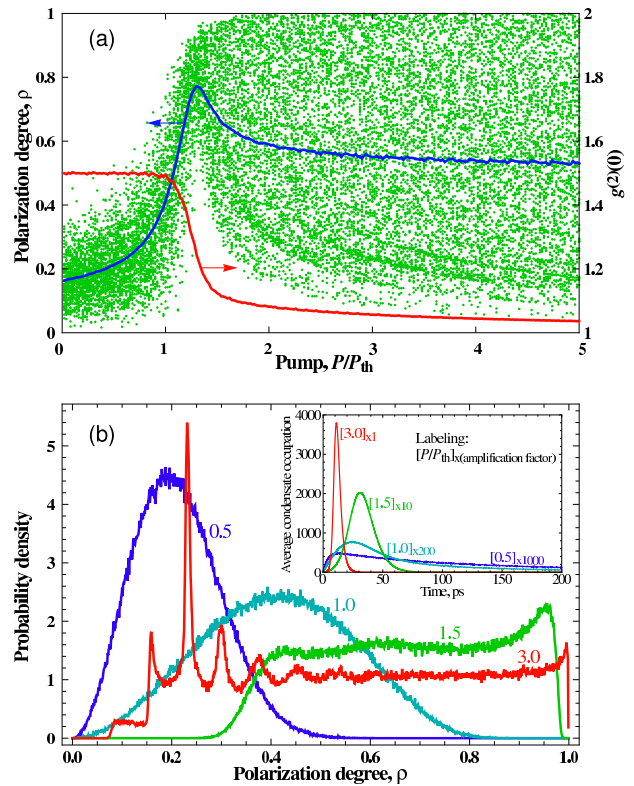}
\caption{\label{Fig1-PolDegree} (a) The total polarization degree
$\rho$ of polariton condensate as a function of pump intensity. Dots show
the random values of $\rho$ for different pulses, while the average
$\left<\rho\right>$ is shown by the solid line. The other solid line
shows the second-order coherence (see text). (b) The distribution
function of the total polarization degree. Curves are labeled by the
values of $P/P_{th}$. The time dependence of the average condensate
occupation number $\left<n\right>$ is shown in the insert for the same
values of $P/P_{th}$.}
\end{figure}

The total polarization degree of the condensate,
\begin{equation}
 \label{TotRho}
 \rho=\frac{2}{\overline{n}}
 \left[
 \left(\overline{S_x}\right)^2+\left(\overline{S_y}\right)^2+\left(\overline{S_z}\right)^2
 \right]^{1/2},
\end{equation}
is shown in Fig.~\ref{Fig1-PolDegree}. One can see that the
solutions to Eqs.~(\ref{LangevinEq}) and (\ref{Reservoir}) exhibit a
well pronounced threshold behavior. The dynamical
threshold~\cite{Rubo03} is defined by the balance of the polariton
income and outcome rates for the condensate:
$W(0)=rN_r(0)=\Gamma_c$, so that the threshold pump is defined by
$\int P_{th} dt=\Gamma_c/r$.

Below threshold, the average occupation of the condensate is less
than 1 and the order parameter fluctuates extensively during the
reservoir lifetime $\Gamma_r^{-1}$ that defines the duration of the
PL signal in this case. The pseudospin also fluctuates strongly both
in amplitude and direction, so that the total polarization degree is
close to zero. It can be shown that $\left<\rho\right>\sim
(\Gamma_r/\Gamma_c)^{1/2}$ for $\Gamma_r\ll\Gamma_c$ and $P\ll
P_{th}$.

Above threshold the condensate is formed during the formation time
$t_f$ such that $W(t)\geqslant\Gamma_c$ for $0<t\leqslant t_f$, and
disappears afterwards on the scale of $\Gamma_c^{-1}$. This leads to
a drastic increase in the condensate occupation and narrowing of the
emission [see insert in Fig.~\ref{Fig1-PolDegree}(b)]. At the same
time there is a strong increase in the total polarization degree of
the condensate. Without polariton-polariton interactions the
polarization degree would reach 1. However, the polariton-polariton
interactions [the third term in Eq.~(\ref{LangevinEq})] result in a
slow decrease of the average polarization degree for $P>P_{th}$. As
a result the dependence $\left<\rho\right>$ on $P$ exhibits a
maximum placed close to the threshold. The build-up of the total
polarization degree indicates the formation of a well-defined order
parameter and well-defined pseudospin for each excitation pulse. It
should be noted, however, that the direction of the pseudospin
randomly changes from pulse to pulse.

Formation of the order parameter at $P>P_{th}$ can be observed also
by measuring the second-order coherence in Hanbury-Brown and Twiss
experiment~\cite{HBTwiss56,Deng02,LeSiDang2008}. The second-order
coherence parameter
$g^{(2)}(0)=\overline{\left<n^2\right>}/\overline{\left<n\right>^2}$
is also shown in Fig.~\ref{Fig1-PolDegree}(a). It is seen that
increase of the pump brings the polariton condensate from the
thermal (chaotic) distribution with $g^{(2)}(0)=3/2$ for $P\ll
P_{th}$~\cite{NoteSOC} to the coherent state with $g^{(2)}(0)\simeq
1$ for $P\gg P_{th}$, where the fluctuations are suppressed and the
order parameter is well-defined for each pulse. Note that the
polariton-polariton interactions that caused a decrease in the
average polarization degree for higher pump powers do not prevent
$g^{(2)}(0)$ from approaching 1. In other words the
polariton-polariton interactions do not result in loss of coherence;
rather they have a dynamical effect on the pseudospin vector, which
affects the average polarization degree as we shall discuss later.
Clearly, both the second-order coherence and polarization
measurements provide evidences of the condensate formation. However,
the polarization measurements give more details about the state of
the condensate since the total polarization is sensitive to the
effects of polariton-polariton interactions.

\begin{figure}[t]
\includegraphics[width=3.2in]{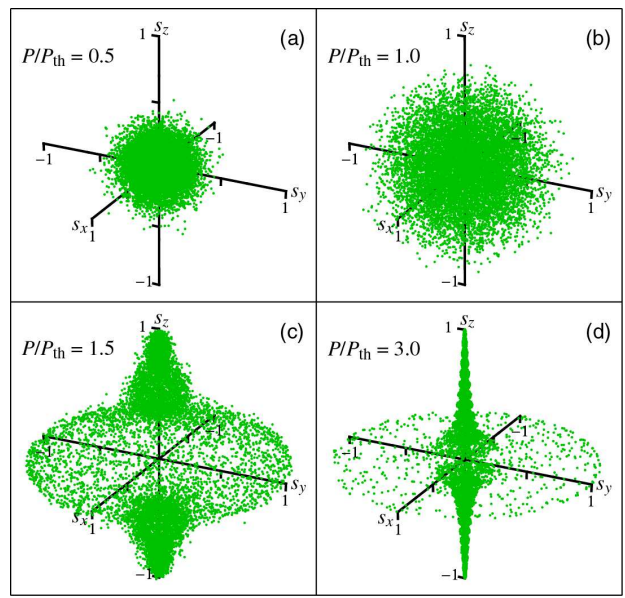}
\caption{\label{Fig2-Rings} Showing the distribution function in
normalized pseudospin space,
$\mathbf{s}=2\overline{\mathbf{S}}/\overline{n}$, for different
values of the excitation pump. Far above the threshold the formation
of the Larmor rings is seen. }
\end{figure}

Another important effect of polariton-polariton interaction is the
nonuniform distribution of the total polarization degree for
$P>P_{th}$. Apart from strong fluctuations from 0 to 1 present in
this case the distribution of polarization degree exhibits sharp and
approximately equidistant peaks that appear far above threshold as
shown in Fig.~\ref{Fig1-PolDegree}(b). The origin of these peaks can
be understood if one examines the condensate polariton distribution
function in the pseudospin space. This distribution function is
shown in Fig.~\ref{Fig2-Rings} for four values of the pump. One can
see that the in-plane component of pseudospin $\sqrt{S_x^2+S_y^2}$
is strongly suppressed for several values of $S_z$ (see
Fig.~\ref{Fig2-Rings}(d) for $P=3P_{th}$). This happens as a result
of spin-anisotropy of polariton-polariton interactions. Indeed, the
interaction Hamiltonian (\ref{Hint}) can be re-written as
$H_{int}=(1/4)(\alpha_1+\alpha_2)n^2+(\alpha_1-\alpha_2)S_z^2$, and
the third term in (\ref{LangevinEq}) produces the self-induced
Larmor precession~\cite{Shelykh2004} of pseudospin around
$\hat{z}$-axis. The frequency of this precession is proportional to
$S_z$. For a fixed $n$ the in-plane component of pseudospin is
averaged to zero for the values of $S_z$ that give full $2\pi$
rotation. In reality the condensate occupation is not fixed and
fluctuations also exist, so that the in-plane pseudospin never
averages to zero exactly. Nevertheless this effect gives rise to a
sequence of ``Larmor'' rings seen at strong pumping.

%%%%%%%%%%%%%%%%%%%%%%%%%%%%%%%%%%%%%%%%%%%%%%%%%%%%%%%%%%%%%%%%%%%%%
\emph{Relaxation of the order parameter.}---The results obtained
above are valid for the case of short condensate lifetime
$\Gamma_c^{-1}$ compared to the typical relaxation times of the
order parameter. In this subsection we show that fast relaxation
modifies qualitatively the polarization properties of the polariton
condensate. Relaxation not only smears out the Larmor rings
discussed above but also favors a linear polarization of the
condensate.

Below we consider the simplest form of the relaxation term, namely,
model A also referred to as the Landau-Khalatnikov or Onsager model
\cite{HHreview77}. In this model, the relaxation term in
Eq.~(\ref{LangevinEq}) is written as $\mathfrak{R}_\sigma=\delta
H_{int}/\delta\psi_\sigma^*$. This term favors relaxation of the
order parameter to minimize the polariton-polariton repulsion
energy. This repulsion energy has its minimum if half of the
polaritons in the condensate have their spins oriented up, whilst
the other half have their spins down, corresponding to a suppression
of the z component of the pseudospin vector and the formation of a
linearly polarized condensate. It should be noted that the
relaxation term we consider does not conserve the number of
polaritons in the condensate. In this way, the condensate depletion
and leakage of polaritons from the condensate are implied by the
present model.

\begin{figure}[t]
\includegraphics[width=3.2in]{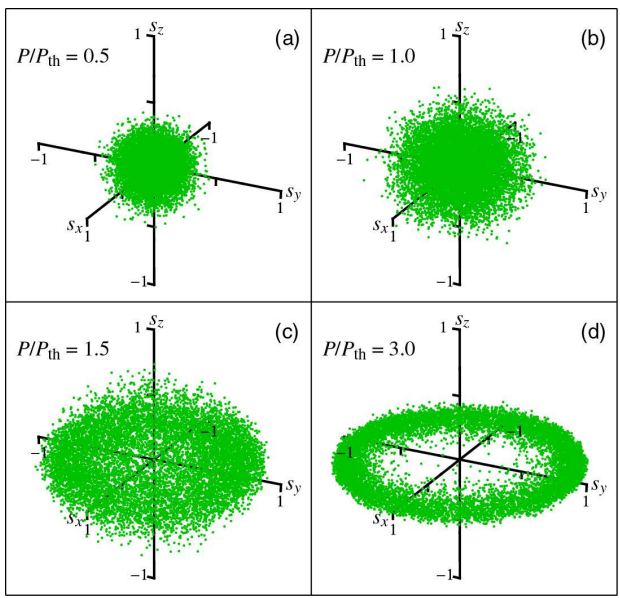}
\caption{\label{Fig3-Torus} Showing the effect of relaxation of the
order parameter on distribution function in normalized pseudospin
space, $\mathbf{s}=2\overline{\mathbf{S}}/\overline{n}$. Contrary to
the case without relaxation (Fig.~\ref{Fig2-Rings}) a linear
polarization is formed for strong pump above threshold. The
relaxation parameter
$\gamma=1\;\mathrm{meV}^{-1}\;\mathrm{ps}^{-1}$. }
\end{figure}

The distribution functions of polaritons in the normalized
pseudospin space for different pump intensities are shown in
Fig.~\ref{Fig3-Torus}. One can see that the circular polarization is
rapidly lost by the condensate as the pumping intensity increases.
The characteristic oscillations of the average linear polarization
of the condensate as a function of its circular polarization degree
seen in Fig.~\ref{Fig2-Rings} vanish once the relaxation of the
order parameter is switched on. The distribution of polariton
condensates in the pseudospin space changes from a sphere to torus
and further approaches a flat ring shape as the pumping power
increases. The condensate develops a strong spontaneous linear
polarization at strong pumping. The build-up of the linear
polarization as a result of the polariton BEC has been
experimentally observed~\cite{Kasprzak06,Balili07}, while in these
works the orientation of linear polarization plane has been pinned
to one of the crystal axes of the structure due to some intrinsic
anisotropy of the samples~\cite{Kasprzak06}. Further experimental
research is needed in order to evidence the stochastic vector
polarization of a polariton condensate.

\emph{Conclusion.}---We have described theoretically the spontaneous
build-up of the vector order parameter in the course of polariton
BEC. The build-up of the order parameter manifests itself in
formation of the stochastic vector polarization of the condensate.
This polarization is correlated with the second order coherence of
the condensate. Its statistical distribution depends on the
mechanism of polarization relaxation in the polariton system.

We acknowledge financial support from the EPSRC. YGR acknowledges
also DGAPA-UNAM for support under the grant No.\ IN107007.

%%%%%%%%%%%%%%%%%%%%%%%%%%%%%%%%%%%%%%%%%%%%%%%%%%%%%%%%%%%%%%%%%%%%%%%%%%%%%%%

%%%%%%%%%%%%%%%%%%%%%%%%%%%%%%%%%%%%%%%%%%%%%%%%%%%%%%%%%%%%%%%%%%%%%%%%%%

\end{document}